\documentclass[12pt]{article}
\usepackage{amsmath,amssymb,amsthm,amsxtra,overpic,bbm,bm,epsfig,ulem,multirow}
\usepackage{color,cite}
\usepackage[symbol]{footmisc}
\usepackage{epstopdf}
\usepackage{booktabs}
\usepackage{footnote}
\usepackage{rotating}
\usepackage{hyperref}
\usepackage{url}
\usepackage{cite}
\usepackage{tikz}
\usetikzlibrary{arrows,shapes,snakes,automata,backgrounds,petri}
\usetikzlibrary{decorations.pathmorphing}
\usetikzlibrary{decorations.markings}
\usepackage{lineno}

\def\thefootnote{\fnsymbol{footnote}}

\usepackage{array}

\vspace{0.2cm}

\begin{document}

\begin{center}
{\Large\bf Pulse shape discrimination technique for diffuse supernova neutrino background search with JUNO} 
\end{center}

\vspace{0.2cm}

\begin{center}
{\bf Jie Cheng~$^{a}$}, 
{\bf Xiao-Jie Luo~$^{b,~c}$}, 
{\bf Gao-Song Li~$^{b}$}~\footnote{Email: ligs@ihep.ac.cn},
{\bf Yu-Feng Li~$^{b,~c}$}~\footnote{Email: liyufeng@ihep.ac.cn},
{\bf Ze-Peng Li~$^{b}$}~\footnote{Email: zel032@physics.ucsd.edu},
{\bf Hao-Qi Lu~$^{b}$},
{\bf Liang-Jian Wen~$^{b}$},
{\bf Michael Wurm~$^{d}$}, and
{\bf Yi-Yu Zhang~$^{b}$},\\

\vspace{0.2cm}
{$^a$School of Nuclear Science and Engineering, North China Electric Power University, Beijing 102206, China}\\
{$^b$Institute of High Energy Physics, Chinese Academy of Sciences, Beijing 100049, China}\\
{$^c$School of Physical Sciences, University of Chinese Academy of Sciences, Beijing 100049, China}\\
{$^d$Institute of Physics and EC PRISMA+, Johannes Gutenberg Universit$\rm \ddot{a}$t Mainz, Mainz, Germany}\\
\end{center}

\vspace{1.5cm}

\date{\today}
\begin{abstract}

Pulse shape discrimination (PSD) is widely used in particle and nuclear physics. Specifically in liquid scintillator detectors, PSD facilitates the classification of different particle types based on their energy deposition patterns. This technique is particularly valuable for studies of the Diffuse Supernova Neutrino Background (DSNB), nucleon decay, and dark matter searches.
This paper presents a detailed investigation of the PSD technique, applied in the DSNB search performed with the Jiangmen Underground Neutrino Observatory (JUNO). Instead of using conventional cut-and-count methods, we employ methods based on Boosted Decision Trees and Neural Networks and compare their capability to distinguish the DSNB signals from the atmospheric neutrino neutral-current background events. The two methods demonstrate comparable performance, resulting in a 50\% to 80\% improvement in signal efficiency compared to a previous study performed for JUNO~\cite{JUNO:2015zny}. Moreover, we study the dependence of the PSD performance on the visible energy and final state composition of the events and find a significant dependence on the presence/absence of $^{11}$C. Finally, we evaluate the impact of the detector effects (photon propagation, PMT dark noise, and waveform reconstruction) on the PSD performance.

\end{abstract}

\def\thefootnote{\arabic{footnote}}
\setcounter{footnote}{0}

\newpage


\section{Introduction}
\label{sec:introduction}

The diffuse supernova neutrino background (DSNB) is a low-energy steady neutrino flux that permeates the universe and originates from the cumulative emissions of all past core collapse supernovae~\cite{Beacom:2010kk,Lunardini:2010ab,Li:2022myd}.
The study of DSNB neutrinos holds great promise for advancing our understanding of the evolution of the universe, core-collapse SNe and the fundamental properties of neutrinos.
The detection of DSNB neutrinos is challenging due to their low flux and energies and the overwhelming background from other neutrino sources. Nevertheless, several experiments, including Super-Kamiokande (SK)~\cite{Super-Kamiokande:2002hei,Super-Kamiokande:2011lwo,Super-Kamiokande:2013ufi,Super-Kamiokande:2021jaq}, KamLAND~\cite{KamLAND:2021gvi, KamLAND:2011bnd} and Borexino~\cite{Borexino:2019wln}, have been improving their search techniques to get a hint of this flux. The next generation, which includes experiments like Jiangmen Underground Neutrino Observatory (JUNO)~\cite{JUNO:2021vlw, JUNO:2022lpc}, Super-Kamiokande Gadolinium experiment (SK-Gd)~\cite{Super-Kamiokande:2023xup} and Hyper-Kamiokande (Hyper-K)~\cite{Hyper-Kamiokande:2018ofw}, will be key for its discovery. 

JUNO will be the largest ever liquid scintillator (LS) experiment and is planned to be in operation for the physical runs by end of 2024~\cite{JUNO:2021vlw}. It is loaded with 20 kton of LS in an acrylic sphere with a diameter of 35.4 m. The DSNB flux can be detected via the inverse beta decay (IBD) channel, especially within the energy window around 10-30 MeV, as the reactor neutrinos will dominate below 10 MeV and the atmospheric neutrino charged current (CC) events will become significant above 30 MeV. The largest background within the energy range of 10-30 MeV comes from the atmospheric neutrino neutral current (NC) interactions on $^{12}$C, whose event rate in the whole detector volume is around 60 per year, in comparison with a rate of 2-3 events per year for the DSNB signal. Therefore, powerful background suppression techniques are required to further increase the signal-to-background ratio, and obtain a significant sensitivity for the DSNB observation.

The pulse shape discrimination (PSD) technique is an excellent method used in LS detectors to distinguish different types of particles based on their energy deposition patterns. This technique relies on the fact that different types of particles produce distinct time profiles of the scintillation light in the LS. Therefore, a precise determination of the scintillation emission time for different particles is essential for the PSD performance. 
In this work, we present a detailed study on the PSD technique that has been used in the context of DSNB sensitivity studies in Ref.~\cite{JUNO:2022lpc}. 

Initially, we employ methods from Ref.~\cite{JUNO:2022lpc,Cheng:2020aaw} to predict the energy spectrum distribution of final-state particles for the DSNB signals and atmospheric neutrino NC backgrounds, which serve as inputs for the detector simulation. Subsequently, we perform a comprehensive simulation using the JUNO offline software framework~\cite{Huang:2017dkh,Lin:2022htc}, incorporating all detector and electronics effects, as well as waveform and event reconstruction. Various factors, including the LS optical model, photon arrival time resolution, and PMT dark noise, are considered in the simulation. Following this, we conduct a systematic analysis of the characteristic photon emission time spectra of positron and non-positron events, extracting them as inputs for different PSD methods. 
Instead of more conventional cut-and-count methods such as the tail-to-total method~\cite{Mollenberg:2014pwa} and the Gatti method~\cite{Gatti},
two PSD methods based on a Boosted Decision Tree (BDT) and a Neural Network (NN), respectively, are used for background suppression. 
Additionally, we evaluate the energy dependence of the PSD performance and, for the first time, evaluate the corresponding efficiency to the DSNB search in JUNO. Finally, we perform a comprehensive investigation of the limiting factors influencing the PSD performance in JUNO by incrementally introducing these factors. 

The paper is organized as follows: the DSNB signals and atmospheric neutrino NC backgrounds are reviewed in Sec.~\ref{sec:2}; the pulse shape discrimination methods are introduced in Sec.~\ref{sec:psd}; then, the performances of both methods are presented and discussed in Sec.~\ref{sec:result}; a summary will be presented in Sec.~\ref{sec:conclusion}.

\section{Signals and backgrounds}
\label{sec:2}

\subsection{DSNB signals}
\label{sec:dsnb_signal}

The DSNB signal prediction, based on Ref.~\cite{JUNO:2022lpc}, depends on a variety of crucial ingredients~\cite{Priya:2017bmm,Kresse:2020nto,Horiuchi:2020jnc}. The first one, which serves as a link to the cosmic history of star formation, is the cosmological SN rate as a function of the progenitor mass and redshift. The second ingredient is the average energy spectrum of SN neutrinos. The last one is the contribution of the failed SNe~\cite{Priya:2017bmm,Kresse:2020nto}, which may alter the DSNB signal and will feature a hotter neutrino energy spectrum compared to the neutron-star-forming SNe (i.e., successful SNe).
A reference DSNB model is adopted in this paper to explore the performance of the pulse shape discrimination technique as described in details in Sec.~\ref{sec:psd}. 
In the reference DSNB model, the failed SNe fraction ($f_{\rm BH}=$ 0.27), the SN Rate at red-shift $z=0$  ($R_{\rm SN}(0)= 1.0\times 10^{-4} \, \rm yr^{-1} \, Mpc^{-3}$) and the average neutrino energy ($\langle  E \rangle =$ 15 MeV) are used to predict the DSNB flux. For details see Ref.~\cite{JUNO:2022lpc}. 

Thanks to the relatively large cross section, the IBD reaction, $\overline{\nu}_{e}+p\rightarrow e^{+} + n$, is the ideal channel for the detection of the DSNB. The interaction produces a positron prompt signal and a delayed neutron capture signal in the LS. In addition, the time coincidence of the prompt and delayed signals arising from the final-state $e^{+}$ and $n$, respectively, perfectly eliminates single-event backgrounds such as signals from the radioactivity of detector materials, the decays of cosmogenic isotopes and the recoil electrons from solar neutrino interactions. We need to take into account the IBD cross section, the target mass, and the detector response in order to calculate the expected DSNB energy spectrum that would be observed in JUNO. We take the free proton number in the JUNO LS as $7.15\times 10^{31} \, \rm kton^{-1}$~\cite{JUNO:2015zny}, assuming a 20 kton of LS. The differential IBD cross section is taken from Ref.~\cite{Strumia:2003zx}. The visible energy of the DSNB signals in JUNO is expected to be below 100 MeV.

\subsection{Atmospheric neutrino NC backgrounds}
\label{sec:atm_nc}

In the search for DSNB, the relevant background source includes reactor $\bar{\nu}_e$'s, NC and CC interaction of atmospheric neutrinos, fast neutrons and long-lived cosmogenic $\beta n$-emitters $^9$Li and $^8$He. Among these, the dominant backgrounds within the energy window ranging from 10 MeV to 30 MeV are the atmospheric neutrino NC background events. It is essential to have a comprehensive understanding of the NC background in order to develop analysis techniques for its suppression. 

As described in Ref.~\cite{Cheng:2020aaw}, we have used dedicated simulations to predict spectra and rates of background events resulting from atmospheric neutrino NC interactions with $^{12}$C nuclei in the LS. Specifically, we employed a representative nuclear model from the \texttt{GENIE} generator (version 2.12.0)~\cite{Andreopoulos:2009rq} to simulate neutrino-nucleus interactions within the JUNO LS detector.  The \texttt{GENIE} model incorporates a nucleon axial-vector form factor parametrization with an axial mass of $\rm M_{A}=$ 0.99 GeV, determined from deuterium measurements~\cite{Kitagaki:1990vs}. To model the nuclear structure, the \texttt{GENIE} model incorporates the relativistic Fermi gas (RFG) model with "Bodek-Ritchie" modifications~\cite{Bodek:1981wr}.  The package \texttt{TALYS} (1.8)~\cite{Koning:2005ezu} is used to describe $n$'s, $p$'s, $\alpha$'s, $\gamma$-rays and other particles from the deexcitations of the final-state nuclei.

Table~\ref{tab:nc_channels} summarizes the rates of DSNB signals and NC interactions between atmospheric neutrinos on $^{12}$C in inclusive and exclusive channels.
For the DSNB search, NC interaction channels with single neutron emission are the major background as they can form prompt-delay coincidences that mimic IBD signals. 
However, while the visible energy (number of photoelectrons) detected is in the same range as that of DSNB-induced positrons, the prompt signals are usually induced by a collection of neutrons, protons, alphas and de-excited gammas (c.f. ``Prompt signal" in Table~\ref{tab:nc_channels}).  
The most relevant NC backgrounds are the quasielastic scattering (QEL) interactions, in which only one neutron is produced, causing a prompt signal with energy below 100 MeV. The expected interaction rates are shown as ``Raw" in Table~\ref{tab:nc_channels}.
A set of selection cuts is used for filtering out IBD events from DSNB signals where the visible prompt energy window is between (12, 30) MeV.
After the IBD selection, the typical predicted rate of DSNB signals in the optimized window is around 0.14 kt$^{-1}$ yr$^{-1}$, which is about one order of magnitude smaller than the atmospheric NC background rate 2.56 kt$^{-1}$ yr$^{-1}$. Therefore, powerful background suppression techniques for non-positron prompt events are necessary to achieve an unambiguous discovery of the DSNB signals. 

\begin{table}[]
    \centering
    \small
    \begin{tabular}{lccccc}
    \hline\hline
                    & & \multicolumn{2}{c}{Raw} & \multicolumn{2}{c}{After IBD selection}\\
    Channel         & Prompt signal & Rate [yr$^{-1}$ kt$^{-1}$] & Fraction [\%] & Rate [yr$^{-1}$ kt$^{-1}$] & Fraction [\%]\\
    \hline\hline
    \multicolumn{6}{c} {\it DSNB} \\
    \hline
    IBD       & $e^{+}$     &  0.16    & 100    & 0.14  & 100 \\
    \hline\hline
    \multicolumn{6}{c}{\it NC interactions of $\nu_{x}+{\rm ^{12}C}\rightarrow \nu_{x} + {\rm X}$ } \\
    \hline
    $n+^{11}$C      & $n$               &2.73   & 28.4    & 0.78  & 30.4 \\
    $n+p+^{10}$B    & $n, p, \gamma$  &3.02   & 31.3    & 0.58  & 22.5 \\
    $n+2p+^{9}$Be   & $n, p, \gamma$  &0.81    & 8.45   & 0.30  & 11.8 \\
    $n+p+\alpha+^{6}$Li  & $n, p, \alpha, \gamma$  &0.67 &6.93 &0.21 &8.30 \\
    $n+p+d+^{8}$Be  & $n, p, d, \gamma$  &0.63 &6.54 &0.21 &8.30 \\
    $n$+others        & $n, x$  & 1.76  & 18.3  & 0.48 & 19.5\\
    \hline
    Total           & $n, x$  & 9.63  & 100   & 2.56 & 100 \\
    \hline\hline
    \end{tabular}
    \caption{Summary of DSNB signal rates and NC interactions of atmospheric neutrino on $^{12}$C in the JUNO sintillator. The NC interactions are listed distinguishing final states, only listing those with neutrons.}
    \label{tab:nc_channels}
\end{table}

\section{Background discrimination with pulse shape discrimination}
\label{sec:psd}

The prompt signal in DSNB IBD events is a $e^+$ accompanied by its annihilation and the consequent emission of two $\gamma$'s (0.511 MeV each),  compared to the complicated composition of $\gamma$, $n$, $\alpha$, and $p$ in NC background events. The intrinsic scintillation timing for $\gamma/e^\pm$ is different from that for $\alpha/p/n$ as the more heavily ionizing particles produce a higher fraction of slow scintillation photons~\cite{Ranucci:1998bc}. In addition, the spatial topologies of energy deposition of $\gamma$'s and $\alpha/p/n$ are different and introduce a different level of broadening effects on the timing spectrum, which can help particle identification. Therefore, the timing information is used to develop PSD methods to discriminate the DSNB signals and NC backgrounds. Detailed descriptions on their implementations are given below. 

\subsection{MC dataset and software framework}

The official JUNO offline software~\cite{Huang:2017dkh} is employed to simulate the DSNB signals and atmospheric neutrino NC backgrounds. The simulation process includes event generation, detector simulation, electronics simulation, waveform reconstruction, and event reconstruction. Below, we provide detailed information on the configurations and procedures involved in each step of the simulation chain.
\begin{description}
    \item[Event generator:] The event generator primarily uses the DSNB spectrum described in Sec.\ref{sec:dsnb_signal}. Additionally, the final states of atmospheric neutrino NC background events, detailed in Sec.\ref{sec:atm_nc}, are also computed and incorporated. Both data samples are uniformly generated within the JUNO LS volume.
    
    \item[Detector simulation:] We perform a comprehensive simulation of the detector utilizing the most up-to-date design geometry~\cite{JUNO:2021vlw}.  
    The energy deposition of particles is simulated using \texttt{Geant4}~\cite{GEANT4:2002zbu} (version 4.10.p02), while the properties of the LS~\cite{Zhang:2020mqz,Ding:2015sys,Buck:2015jxa} and scintillation photon emission time profiles\cite{neu2020Stock} are based on experimental measurements. Optical processes, including photon absorption, re-emission, scattering, refraction, and total reflection, are implemented for photons propagating in the LS~\cite{JUNO:2020bcl}. Based on measurements of the photon detection efficiency~\cite{Zhang:2020kkr}, 
    for every 1 MeV of deposited energy at the center of the detector, an average of approximately 1350 photo-electrons (p.e.) is detected by the photomultiplier tubes (PMTs).
    A ``hit" refers to the p.e. detection by a PMT. In this study, all hits collected by the PMTs, not just the first one, are utilized.
    
    \item[Electronics simulation:]
    The electronics response is modeled in the simulation including the transition time spread (TTS) and the dark noise rate of the PMTs. 
    In JUNO, two types of high quantum efficiency 20-$\text{inch}$ PMTs~\cite{Wang:2017xvc} are used, namely Hamamatsu dynode PMTs and NNVT PMTs with a novel Micro-channel Plate (MCP) design. The total number of MCP PMTs and dynode PMTs is 12,612 and 5,000, respectively, providing 75\% photocathode coverage. The TTS values for MCP PMTs and dynode PMTs are 12 ns and 2.8 ns, respectively, with dynode PMTs offering significantly better time resolution. The dark noise rate for MCP PMTs is typically around 30 kHz, which is approximately twice as high as that of dynode PMTs. 25,600 3-$\text{inch}$ PMTs are installed in the gaps between the 20-$\text{inch}$ PMTs. However, their data is not utilized in this study due to a low photo coverage of only 3\%~\cite{Li:2018fny}.
     
    \item[Reconstruction:]  
    The waveforms of the PMTs are generated using realistic measurements of single p.e. (SPE) shapes. These waveforms are then digitized and stored for offline data reconstruction. During this process, the time and charge information of PMT hits is reconstructed. Furthermore, a reconstruction algorithm~\cite{Huang:2022zum} is applied to estimate the total event energy and vertex, leading to in a reconstructed vertex resolution that is greater than 15 cm. 
\end{description}

\begin{figure}
    \centering
    \includegraphics[width=0.6\linewidth]{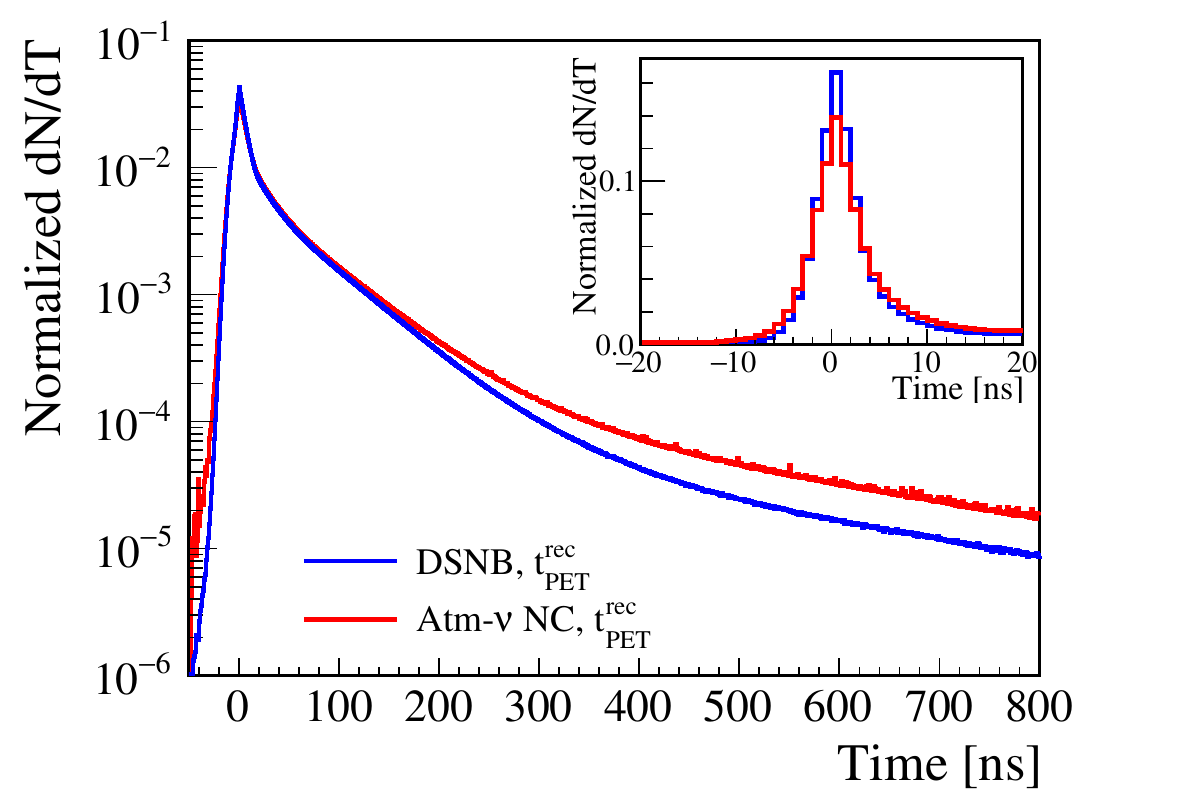}
    \caption{The average reconstructed photon emission time for the DSNB signals (blue) and the atmospheric neutrino NC backgrounds (red). The peak structures of both the DSNB signal and the NC background events are zoomed in on the sub-panel. The dark noise is subtracted from the distributions for demonstration purposes.}
    \label{fig:pet}
\end{figure}

The time-of-flight (TOF) is calculated using the reconstructed vertex, assuming an optical path that follows a straight line and taking into account the effective refractive indexes as described in Ref~\cite{Li:2021oos}.
The photon emission time (PET) is reconstructed by subtracting the TOF from the hit time obtained through the waveform reconstruction. 
It should be noted that waveform reconstruction can only determine the arrival time of the first photon when multiple single-photon waveforms overlap in a PMT within 1 ns. To mitigate the impact of this effect on the time profile, the hit time is weighted by the charge for the subsequent PSD methods.
In Fig.~\ref{fig:pet}, we present the average normalized reconstructed PET distributions for both the DSNB signal and the NC background events. The peak time for each event is aligned to 0 ns. Importantly, the NC event distributions display elongated tails in contrast to the DSNB signal distributions. Additionally, the peaks in the NC event distributions exhibit a relatively shallower shape compared to the DSNB signal distributions. This difference, that will be exploited by the PSD techniques, can be ascribed to the extended travel distance of neutron recoils within the prompt signal of NC events. 
The PSD algorithms are specifically designed to differentiate between signal and background events based on their timing spectral shape differences. However, several factors can adversely impact the performance of PSD. Further discussion on these factors will be presented in Sec.\ref{sec:limit_discussion}.

\subsection{Boosted Decision Tree (BDT) method}
\label{sec:bdt}
 
\begin{figure}[!tb]
    \centering
    \includegraphics[width=1.0\linewidth]{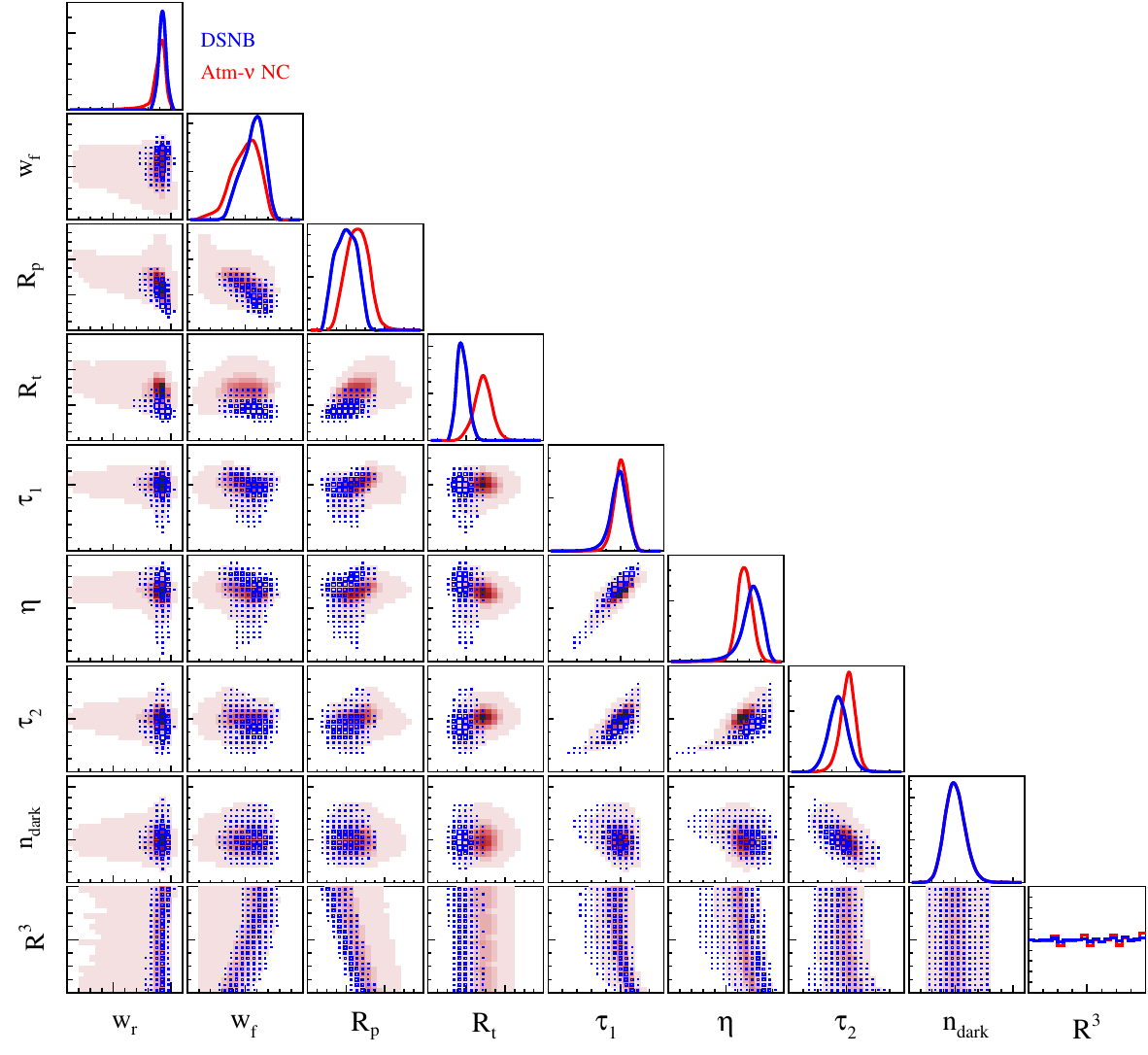}
    \caption{The area normalized distributions of the input variables for the BDT method from the DSNB signals (blue curve) and atmospheric neutrino NC backgrounds (red curve) have been presented in the diagonal panels. In the lower panels, the scatter matrix of the correlations between the different input variables is shown, with the DSNB signals and atmospheric neutrino NC backgrounds denoted by blue and red dots, respectively.}
    \label{fig:bdt_input_vars}
\end{figure}

In our study, we have employed the BDT method~\cite{Roe:2004na} for particle identification, recognizing that multivariate machine learning techniques offer a promising alternative to conventional cut-and-count methods~\cite{Mollenberg:2014pwa}. Tab.~\ref{tab:input_vars} provides a summary of the input variables used in the BDT method, based on the features observed in time profiles of different particles. 
Fig.~\ref{fig:bdt_input_vars} displays the input variable distributions along with the correlation matrix among these variables.
The definitions of the variables are explained below.
\begin{table}[]
    \setlength{\tabcolsep}{9pt}
    \centering
    \begin{tabular}{c|c|c}
    \hline
             Variable & Unit & Physics meaning  \\
    \hline
    \hline
    $w_\mathrm{r}$ &ns   & Peak rising time within [-20, 20] ns \\
    $w_\mathrm{f}$ &ns   & Peak falling time within [-20, 20] ns\\
    $R_\mathrm{p}$ &- & Fraction of the pulse area in [20, 1100] ns \\
    $R_\mathrm{t}$ &- & Fraction of the pulse area in [400, 1100] ns \\
    $\tau_1$ &ns & Decay time of the fast component of time spectrum in [40, 1100] ns \\
    $\tau_2$ &ns & Decay time of the slow component of time spectrum in [40, 1100] ns \\
    $\eta$ &- & Fraction of the fast component of time spectrum in [40, 1100] ns \\
    $n_\mathrm{dark}$ &ns$^{-1}$ & Dark noise rate \\
    $R^3$ & m$^{3}$ & Cubic of radius of the reconstructed event position\\
    \hline
    \end{tabular}
    \vspace{0.3cm}
    \caption{A set of input variables for the BDT method.}
    \label{tab:input_vars}
\end{table}
\begin{itemize}
 \item $w_\mathrm{r}$ and $w_\mathrm{f}$ respectively denote the rising and falling times within the interval [-20, 20] ns. These ascending and descending segments within the waveforms contribute to 10\% of the total area encompassed by this time range. These two variables effectively capture the distinctive features found in the peak regions of the temporal profiles. The calculations for $w_\mathrm{r}$ and $w_\mathrm{f}$ are based on hits from the dynode 20-inch PMTs. The utilization of these specific PMTs preserves intrinsic peak profile characteristics, facilitated by their better time resolution. 
 
  \item $R_\mathrm{p}$ and $R_\mathrm{t}$ are defined as the fractions of the pulse area within [20, 1100] ns and [400, 1100] ns, respectively. These two variables are used to further distinguish peak shape and tail shape of different particles, and are also calculated using hits from dynode PMTs.
 
  \item The tail distributions of the time spectra follow a multi-exponential probability density function (PDF). To model this distribution, we have developed a two-exponential fit as shown in Eq.(\ref{eq:fit}), using the un-binned maximum-likelihood method. The fit is applied to the time spectra within the range of [40, 1100] ns.
  
  \begin{equation}
  f(t) = N\times\left[\frac{\eta}{\tau_{1}}\,{\rm exp}\left(-\frac{t}{\tau_{1}}\right)+\frac{(1-\eta)}{\tau_{2}}\,{\rm exp}\left(-\frac{t}{\tau_{2}}\right)\right] + n_\mathrm{dark}
  \label{eq:fit}
  \end{equation}

    The exponential terms in Eq.(\ref{eq:fit}) represent the fast and slow components of the time spectrum. The number of hits associated with these components is denoted as $N$, with corresponding decay times $\tau_{1}$ and $\tau_{2}$. The parameter $\eta$ represents the fraction of the fast component within the time spectrum range of [40, 1100] ns. The recorded time spectrum includes dark noise hits from PMTs that can not be distinguished from physics hits. In order to accurately estimate the dark noise rate ($n_\mathrm{dark}$), a fitting parameter is introduced, using the time spectrum within the range of [-200, -100] ns, where dark noise hits are present. Given the restricted statistical dataset, the fit value of $n_\mathrm{dark}$ displays correlation with other input variables, particularly the effective slow component $\tau_2$. Nonetheless, the incorporation of $n_\mathrm{dark}$ contributes to a more accurate representation of the time spectrum. To improve the statistical precision, all hits from both dynode and MCP PMTs are used in the fitted time profile.

  \item $R^3$ is defined as the radius cubed of the reconstructed event position, $R$ being the distance from the reconstructed interaction vertex to the center of the detector. This parameter is also used as an input variable in the BDT method to mitigate the impact of the TOF bias on the PSD performance. As shown in Fig.~\ref{fig:bdt_input_vars}, it is evident that the variables associated with the peak area or those in close proximity to the peak area exhibit correlations with the positions. Hence, the variable $R^3$ is included to enhance the discrimination capabilities of $w_\mathrm{f}$, $R_\mathrm{p}$, $\tau_1$, and $\eta$. 

\end{itemize}

The BDT method, integrated into the Toolkit for Multivariate Analysis (TMVA\footnote{The data analysis framework ROOT includes the TMVA toolkit, which offers a wide range of multivariate classification algorithms. In this work, the version of ROOT utilized is v6.22/08.})~\cite{TMVA:2007ngy}, is a widely used supervised learning algorithm in particle physics for classification tasks. It involves combining multiple decision trees to create a strong predictive model. Each decision tree makes a series of binary decisions based on input features (discriminating variables, as shown in Tab.~\ref{tab:input_vars}), eventually leading a prediction or classification (e.g., signal or background). 
The BDT model is trained using 1 million signal events and 1 million background events.  
According to the importance ranking report from TMVA, the variables $\eta$, $\tau_{2}$, and $\tau_{1}$ are identified as the most important ones. Additionally, other variables are introduced to further enhance the PSD performances.
The distribution of the test samples is consistent with that of the training samples, indicating no evidence of over-training. The results will be discussed in Sec.~\ref{sec:bdt_performance}.

\subsection{Neural Network (NN) method}

An alternative discriminator has been developed based on NN methods. It is worth noting that a similar NN method has been implemented for $\alpha/\beta$ pulse shape discrimination in Borexino~\cite{BOREXINO:2023pcv}. Instead of extracting features from the event time distributions, the latter are directly used as the input. All the hits in an event are binned in time and an array of the number of hits in each bin is taken as the input. Due to the exponential decay characteristic of LS and the need to gather sufficient statistics within each bin, an uneven binning strategy has been implemented.  The binning scheme incorporates a carefully designed range of bin widths, spanning from finer intervals to broader ones. 

The event time distributions are weighted by the reconstructed charge of each hit to accurately accommodate the influence of multiple p.e. hits, following the same methodology as the BDT method outlined in Section~\ref{sec:bdt}.
To mitigate the broadening effect caused by the relatively broad SPE response of the MCP PMTs, unweighted time distributions are also provided as a supplementary input.
Moreover, to capture potential vertex-dependent attributes, the reconstructed vertex information $R^3$ is appended to the input array. These three components are then concatenated into the one-dimensional input for the network.

\begin{figure}[!tb]
    \centering
    \includegraphics[width=0.9\linewidth]{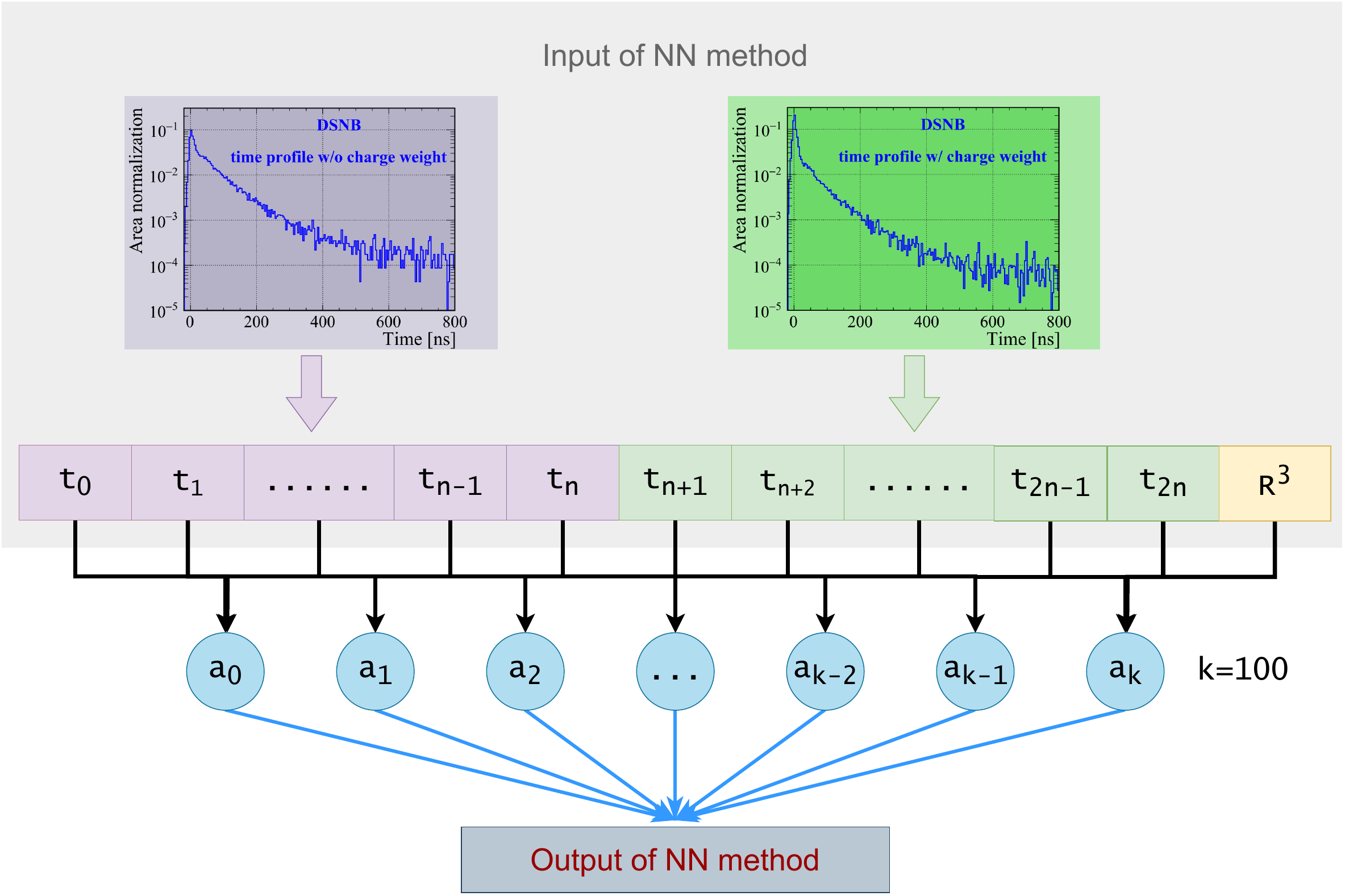}
    \caption{The structure of the neural network. The input is a one-dimensional array encoded with charge and time information of PMT hits and the event vertex $R^3$. The network is a fully connected layer composed of 100 nodes and $a_0$ represent constant term. The output is a signal-like score ranging from 0 to 1.}
    \label{fig:MLP Structure}
\end{figure}

A simple NN with one fully connected layer is chosen for event classification. The network is implemented with the MLPClassifier (multi-layer perceptron classifier) module from the scikit-learn toolkit \cite{Pedregosa:2011ork}. The fully connected layer consists of 100 nodes. An illustration of the input and network structure is shown in Fig.~\ref{fig:MLP Structure}. 
The training dataset matches the one employed in the BDT method. Approximately seven hundred thousand samples of DSNB signals and atmospheric neutrino NC backgrounds are used for training the model, while the remaining events are reserved for testing purposes.

\section{Results and discussions}\label{sec:result}
\subsection{Comparison of BDT and NN methods: PSD performance}
\label{sec:bdt_performance}

\begin{figure}[!tb]
    \centering
    \includegraphics[width=0.9\linewidth]{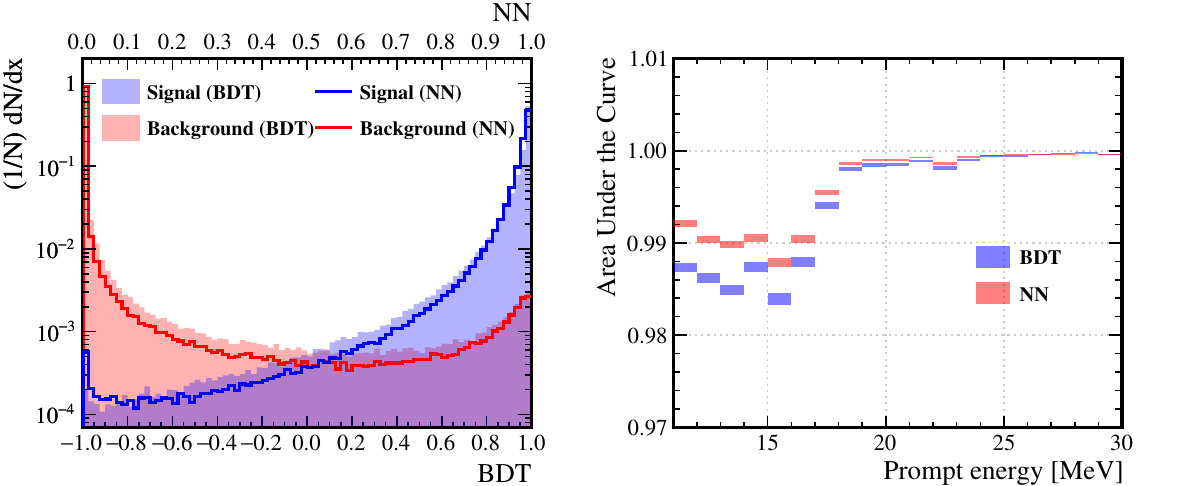}
    \caption{The left panel illustrates the BDT (filled area) and NN (solid lines) distributions. The right panel displays the areas under the ROC curve as a function of the prompt energy from both methods, with the shaded band representing the statistical error.}
    \label{fig:bdt_1d}
\end{figure}

The left panel of Fig.~\ref{fig:bdt_1d} presents the PSD distributions obtained from the BDT and NN methods. Both methods exhibit similar trends in the one-dimensional PSD distributions, demonstrating clear separation between signal and background events.  
In accordance with the requirements of the latest DSNB analysis, we choose 
a residual fraction of 1\% for the NC background. Under this condition, within the prompt energy range of [11, 30] MeV, a signal efficiency of around 80\% is achieved. 
The right panel of Fig.~\ref{fig:bdt_1d} shows the area under the Receiver Operating Characteristic (ROC) curve distributions as a function of the prompt energy for the BDT and NN methods. Both methods exhibit a similar energy-dependent trend, with the NN method demonstrating better PSD performance at energies below 20 MeV.

\begin{figure}[!tb]
    \centering
    \includegraphics[width=0.9\linewidth]{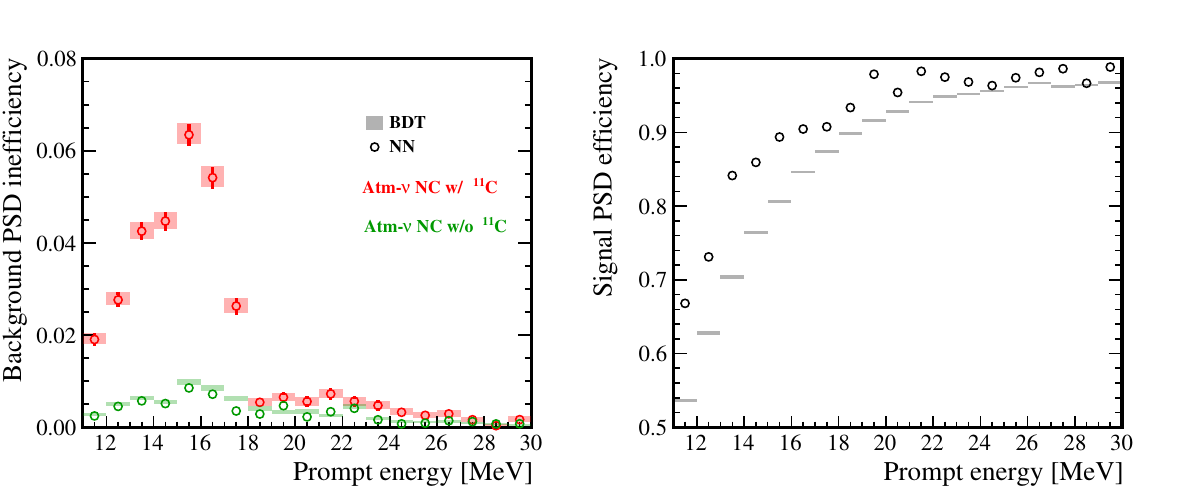}
    \caption{The background inefficiency (left) and signal efficiency (right) as a function of the prompt energy with a PSD cut retaining 1\% total background. The legend in the right plot is identical to the one in the left plot. In each prompt energy bin, the cut on NN discriminator is chosen to achieve the same background inefficiency as BDT method. The shaded band in the BDT and the error bar in the NN represent statistical errors. The BDT and NN methods give similar energy dependence and separation. Atm-$\nu$ NC w/ $^{11}$C events with prompt energy below 18 MeV are more difficult to reject.}
    \label{fig:bdt_efficiency_ene}
\end{figure}

The PSD efficiency is energy-dependent, as illustrated in Fig.~\ref{fig:bdt_efficiency_ene}. The figure includes both BDT and NN. Again, a PSD using either BDT  or NN discriminators is applied, choosing to maintain an NC background fraction of 1\%. The residual NC background events are then categorized into two types based on their final states: one type corresponds to interactions with $^{11}$C in the final state (atm-$\nu$ NC w/ $^{11}$C), the other represents interactions without $^{11}$C (atm-$\nu$ NC w/o $^{11}$C). As expected, the PSD performance improves with rising visible energy due to the large photon statistics available. 
Moreover, both discriminators are more efficient identifying atm-$\nu$ NC w/o $^{11}$C events, since apart from $n$, the prompt energy of these events also originates from $\alpha$, $p$, and other particles (c.f. Tab.~\ref{tab:nc_channels}).
The atm-$\nu$ NC w/ $^{11}$C events are more difficult to distinguish because the single high-energy neutron in the final state can produce gamma rays in inelastic scattering with carbon nuclei~\cite{rinard1991neutron}. Inelastic scattering of high energy neutrons on $^{12}$C excites the nucleus to excited states, with high energy gamma emission from de-excitation. The electron-like gamma contribution to the prompt energy makes it indistinguishable from the DSNB signal events, which explains the feature of increased residual background below 18 MeV for atm-$\nu$ NC w/ $^{11}$C in Fig.~\ref{fig:bdt_efficiency_ene}. More details will be discussed in Sec.~\ref{sec:c11_discussion}.

To facilitate the comparison, varying PSD cuts are implemented on the NN discriminator, ensuring that the PSD inefficiency for the atm-$\nu$ NC w/ $^{11}$C background aligns consistently with the BDT method across different energy bins. Under these conditions, the NN method exhibits a similar pattern for atm-$\nu$ NC w/o $^{11}$C background, but features an improved signal efficiency for the DSNB signal sample. 

\subsection{Energy dependent efficiency for atm-$\nu$ NC with $^{11}$C}
\label{sec:c11_discussion}

\begin{figure}[!tb]
    \centering
    \includegraphics[width=0.9\linewidth]{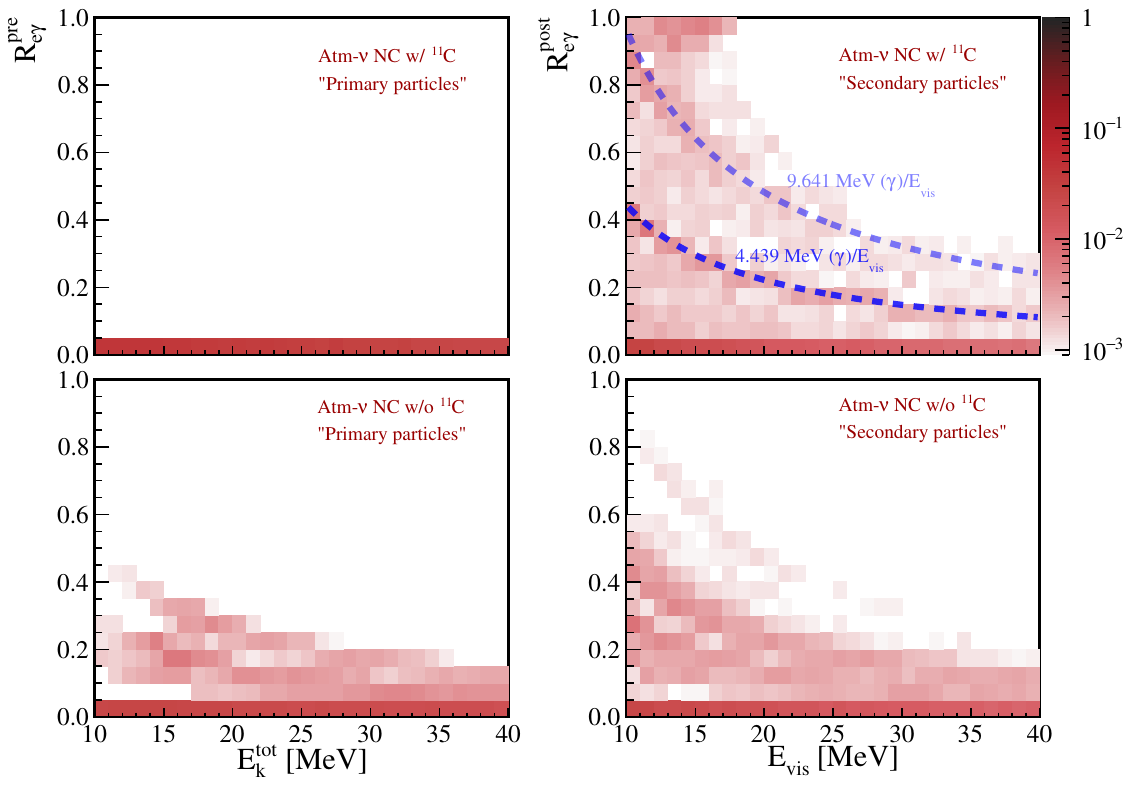}
    \caption{The $R_{e\gamma}$ distributions of the exclusive final state channels for atmospheric neutrino NC background events are depicted as a function of energy in the pre-simulation (left panels) and post-simulation (right panels) scenarios. The total kinetic energy ($E_\mathrm{k}^\mathrm{tot}$) of all final state particles, excluding the neutrino, represents the energy before the detector simulation, whereas the visible energy of the prompt signal ($E_\mathrm{vis}$) represents the energy after the simulation. The exclusive final state channels are categorized into atm-$\nu$ NC w/ $^{11}$C in top panels and atm-$\nu$ NC w/o $^{11}$C in bottom panels. }
    \label{fig:r_eg_vs_others}
    
\end{figure}

A clear energy dependence in the PSD efficiency is observed for the atm-$\nu$ NC w/ $^{11}$C channel. To investigate this, the $e\gamma$ ratio $R_{e\gamma}$ is defined as the fraction of energy contributed by $e^{\pm}$ and $\gamma$ particles over the total visible energy. The prompt energy of the NC background events is a composite of $e^\pm$, $\gamma$, $\alpha$, $p$ signals. When more photons are generated from $e^{\pm}$, $\gamma$, the hit time distributions get closer to those of DSNB signals. Fig.~\ref{fig:r_eg_vs_others} illustrates the distributions of $R_{e\gamma}$ as a function of energy for atmospheric neutrino NC background events with and without $^{11}$C.

The left panels of Fig.~\ref{fig:r_eg_vs_others} display the scenarios involving the primary particles from the initial neutrino interaction, where $R_{e\gamma}^{\rm pre}$ is defined as the ratio of the kinetic energy of $e^{\pm}$ and $\gamma$ particles to the total kinetic energy of all final state particles, excluding the neutrino. According to the deexcitation model used in predicting NC background events, the atm-$\nu$ NC w/ $^{11}$C events have no deexcitation $\gamma$'s that results in $R_{e\gamma}^{\rm pre}$ being equal to 0. Atm-$\nu$ NC w/o $^{11}$C events could have deexcitation $\gamma$'s originating from excited residual nuclei leading to nonzero $R_{e\gamma}^{\rm pre}$, but the majority of events have $R_{e\gamma}^{\rm pre}$ close to 0. 

The secondary particles from subsequent interactions of product particles in the detector, are shown in the right panels of Fig.~\ref{fig:r_eg_vs_others}. The $R_{e\gamma}^{\rm post}$ is defined as the ratio of the visible energy contributed by $e^{\pm}$ and $\gamma$ particles to the total visible energy. This calculation is performed using the MC data obtained in the post-simulation stage. In comparison to the scenarios involving the primary particles, for the atm-$\nu$ NC w/ $^{11}$C events, the distribution of $R_{e\gamma}^{\rm post}$ exhibits two additional bands, which are caused by two types of deexcitation $\gamma$ rays with energies of 4.439 and 9.641 MeV. This phenomenon arises because the energetic fast neutrons from the atm-$\nu$ NC w/ $^{11}$C events can excite the $^{12}$C nucleus to an excited state, and the subsequent deexcitation $\gamma$ emission contributes a significant portion of the emitted photons. Similar event typologies have been observed for neutron interactions with oxygen in water Cherenkov detectors~\cite{Maksimovic:2021dmz,Super-Kamiokande:2019hga}.

\begin{figure}[!tb]
    \centering
    \includegraphics[width=0.7\linewidth]{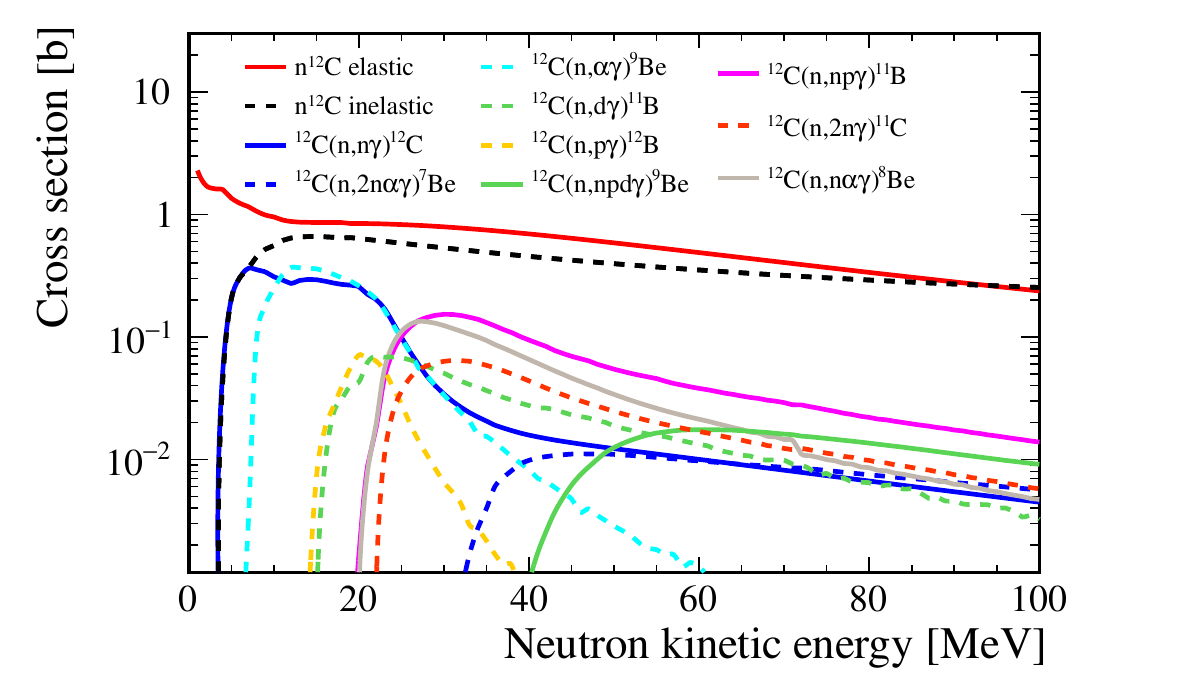}
    \caption{The cross sections of exclusive $n - ^{12}$C reactions, extracted from the results of \texttt{TALYS}, as a function of the incident neutron kinetic energy.}
    \label{fig:nC12xsec}
\end{figure}

To gain a deeper understanding of the differences in $R_{e\gamma}$ before and after the simulation, we use \texttt{TALYS} to calculate the cross sections of exclusive $n - ^{12}$C reactions at various incident neutron energies. These results are depicted in Fig.~\ref{fig:nC12xsec}.
At kinetic energies of the fast neutron below approximately 10 MeV, the dominant interaction between $n$ and $^{12}$C is elastic scattering ($n^{12}$C). However, as the energy exceeds approximately 10 MeV, the inelastic scattering cross section of $n^{12}$C increases significantly, competing with the elastic scattering process. The inelastic process $^{12}$C($n$, $n\gamma$)$^{12}$C dominates the cross section up to around 20 MeV. Subsequently, processes such as $^{12}$C($n$,$np\gamma$)$^{11}$B become important, and the emission of kicked-out $p$ and $\alpha$ particles begins to dominate the photon emission process. This explains the presence of a substantial population at around 15 MeV.

For events with $R_{e\gamma}^{\rm post}<0.6$, they can be effectively identified as background with a PSD score (BDT/NN) close to -1/0 (indicating a background-like event). However, for the remaining events, a positive correlation is observed between $R_{e\gamma}^{\rm post}$ and the PSD score. 

\subsection{Limiting factors for the performance of discriminators}
\label{sec:limit_discussion}

To explore the theoretical limits of separating the DSNB signals from atmospheric neutrino NC backgrounds, we systematically study the PSD capability using the photon emission time distributions obtained from different simulation stages to perform PSD. Detector and reconstruction effects can cause blurring of the reconstructed hit time, leading to a loss of differentiation between the two signal types. Therefore, we divide the complete simulation chain into five different simulation stages and provide detailed comments on these stages to better understand the simulation results:

\begin{description}
    
      \item[Stage 1:] Events are simulated from the beginning at the event generator to the point of initial photon emission. The initial PET value ($t_{\rm PET}^{s1}$) is obtained using detector simulation truth information.
      \item[Stage 2:] This stage corresponds to the complete detector simulation.
      Optical processes of a photon from its primary position to a PMT are simulated, and the true PMT hit time is obtained. The associated PET value ($t_{\rm PET}^{s2}$) is evaluated by subtracting the TOF from the true PMT hit time. The degradation of PSD performance in this stage demonstrates the impact of the calculated vs. the true TOF.  
      \item[Stage 3:] Electronics simulation is performed to incorporate more realistic electronics effects, including TTS. The hit time obtained from this simulation is used to determine the associated PET value ($t_{\rm PET}^{s3}$), which is used to evaluate the impact of TTS on PSD performance. 
      \item[Stage 4 and Stage 5:] Waveform and event reconstructions are conducted. {\bf {Stage 4}} doesn't include dark noise, while {\bf {Stage 5}} includes dark noise. These settings are used to study the effect of reconstruction and dark noise on particle discrimination. The corresponding reconstructed PET values, $t_{\rm PET}^{s4}$ and $t_{\rm PET}^{s5}$, are obtained after the full simulation chain.
      
\end{description}

\begin{figure}[!tb]
    \centering
    \includegraphics[width=1.\linewidth]{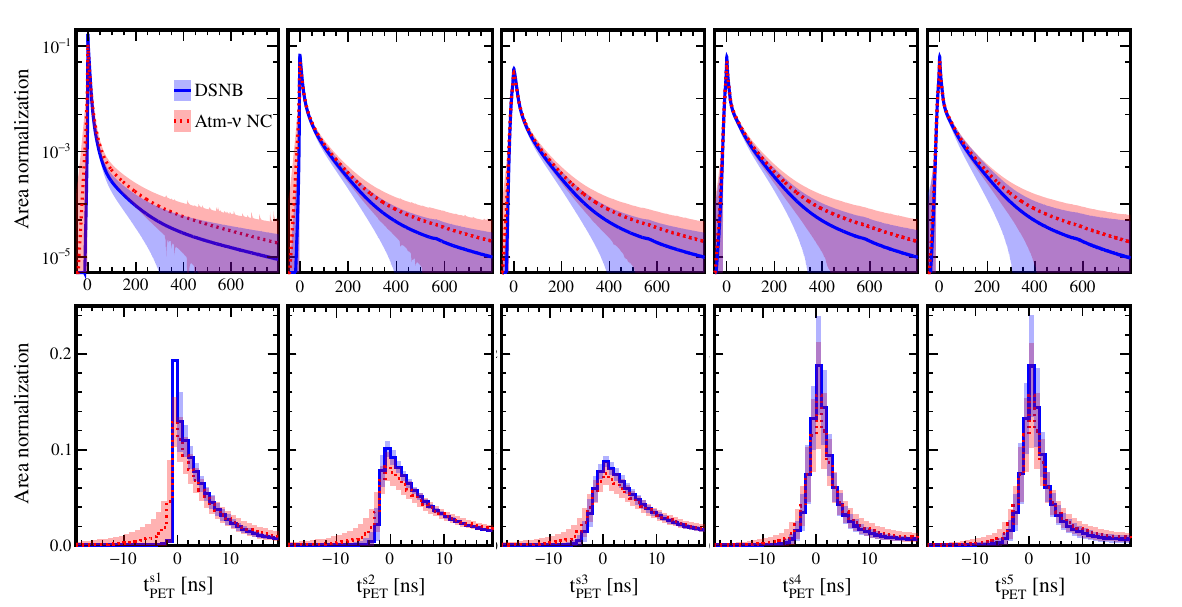}
    \caption{The photon emission time distributions for the DSNB signals (blue) and atmospheric neutrino NC backgrounds (red) have been illustrated at five different simulation stages. Each curve represents the averaged time profile, and the colored band along each curve indicates a 1$\sigma$ deviation from the average. The lower panels display the time profiles around the peak using hits from dynode PMTs, following the same format as the top panels. The two rightmost panels illustrate the impact of dark noise on time profiles.}
    \label{fig:pet_diff_phases}
\end{figure}

Fig.~\ref{fig:pet_diff_phases} shows the photon emission time distributions for the DSNB signals and atmospheric neutrino NC backgrounds at the five simulation stages. The curves in the panels represent the averaged time profile, progressing from left to right as {\bf {Stage $i$}} for $i=$ 1-5. During these stages, the introduction of detector, electronics, and reconstruction impacts causes changes in the average time shape. The shaded band along each curve represents a 1$\sigma$ deviation from the averaged time profile. Shape fluctuations primarily arise from statistical uncertainty.
The time range shown in the upper panels is -50 to 800 ns. The panels offer interesting insights. Firstly, the averaged PET profiles from {\bf {Stage 2}} exhibit a delay between around 50 and 300 ns compared to the true time profiles. Since the actual photon path is unknown, a straight line connecting the event vertex and the fired PMT is assumed, which can introduce a bias towards small values of TOF for scattered or reflected photons. Also, optical refraction on the vessel is neglected. Secondly, the average tail time spectra remain consistent throughout all simulation stages, but their shape fluctuation gradually increases. 

The lower panels focus on the peak shape of the corresponding time profiles, specifically using hits from dynode PMTs. In the lower panels, the time range is zoomed in -20 to 20 ns. The time profiles from {\bf {Stage 1}} illustrate that the prompt recoil neutron of NC events travels a longer distance and takes more time to deposit its energy. This characteristic primarily impacts the distribution of starting times, thereby providing another discriminating factor in the peak time profiles between DSNB signal and NC background events.
As the simulation progresses, the expected blurring of the distinguishing characteristics of this difference is observed in {\bf {Stage 2}} and {\bf {Stage 3}}. However, during the reconstruction stage, a sudden sharpening of the peak shape is noticed. This is attributed to the generation of a higher number of photons at the initial stage within a short time frame. The overlapping of these multiple original time waveforms results in the waveform reconstruction detecting only the time of the first photon and assigning it the combined charge of the closeby p.e.'s. Subsequently, when the PET is obtained from the waveform reconstruction and weighted by charge, the peak becomes sharper. The larger impact on DSNB events actually enhances the differences between peaks to some extent, which is reflected in the performance of the PSD.

\begin{figure}[!tb]
    \centering
    \includegraphics[width=0.6\linewidth]{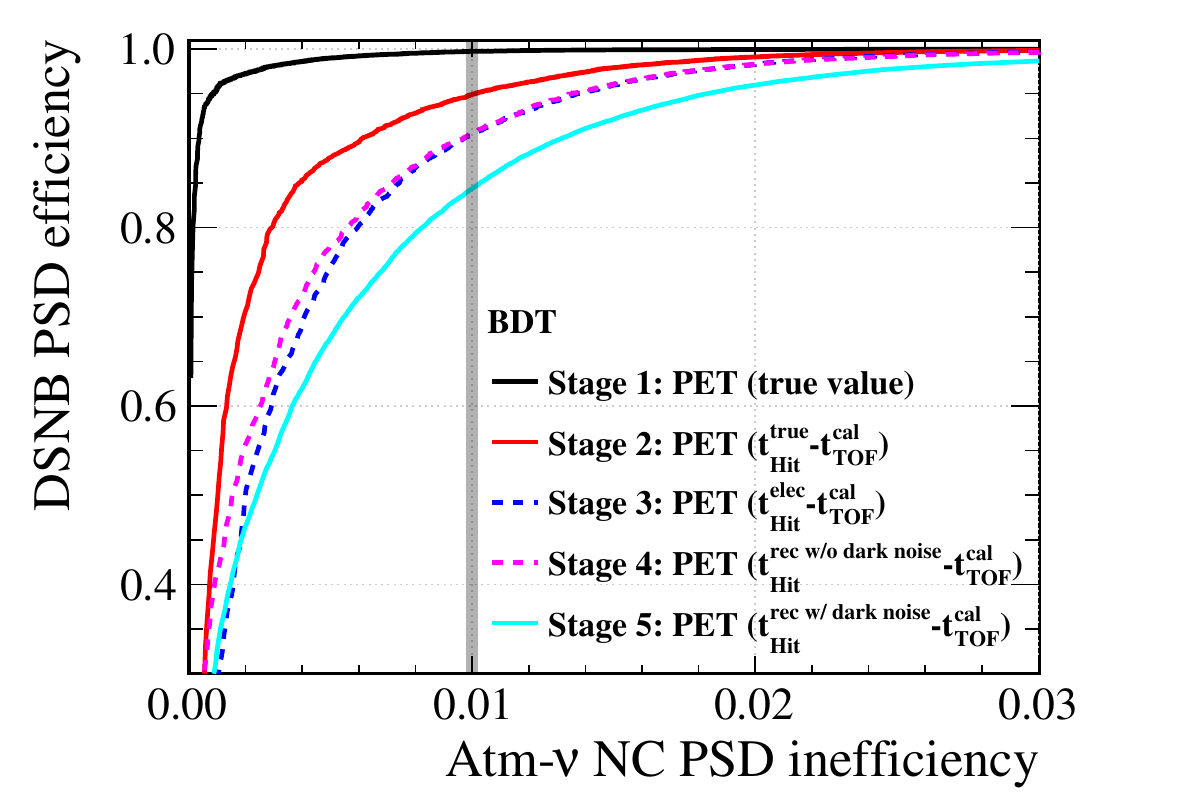}
    \caption{DSNB PSD efficiency as a function of atmospheric neutrino NC background PSD inefficiency using the BDT method in the energy range of 11-30 MeV at five different simulation stages.}
    \label{fig:roc_5stages}
\end{figure}

The BDT method is employed to assess the theoretical limitations of distinguishing the DSNB signals from atmospheric neutrino NC backgrounds. By using the time profiles obtained from these five stages (depicted in Fig.~\ref{fig:pet_diff_phases}) as inputs for the BDT method, the performance of the PSD is evaluated. Note that for the BDT discriminator, $n_\mathrm{dark}$ is fixed to 0 in {\bf {Stages 1-4}}. Fig.~\ref{fig:roc_5stages} illustrates the PSD efficiency for DSNB signals as a function of the  PSD inefficiency for atmospheric neutrino NC background within the energy range of 11-30 MeV at {\bf {Stages $i$}} for $i=$ 1-5. The results indicate that the inclusion of additional detector and electronic effects degrades the PSD performance. However, the PSD performance in {\bf {Stage 4}} is comparable to or slightly better than that in {\bf {Stage 3}}. As mentioned earlier, the sharper peak in the time profiles of {\bf {Stage 4}} provides additional discrimination power. If an average NC background PSD inefficiency of 1\% is required, the DSNB PSD efficiency is approximately 100\% starting from {\bf {Stage 1}}. However, it decreases to 95\% due to the impact of calculated TOF ({\bf {Stage 2}}), further reduces to 90\% due to the uncertainty of electronics effects ({\bf {Stage 3}}), and finally drops to 84\% due to the pollution of dark noise ({\bf {Stage 5}}). 

\section{Conclusion and outlook}
\label{sec:conclusion}
In this paper, we presented the details and understanding of the PSD variables developed for suppressing atmospheric neutrino NC backgrounds, dominant for the DSNB search in JUNO. The particle-type dependent scintillation time profiles of liquid scintillators allow for an efficient particle identification. Using state-of-the-art parameters of the JUNO LS recipe~\cite{neu2020Stock}, we developed two PSD discriminators based on machine learning methods. The BDT and NN methods give consistent results, with the NN method exhibiting superior PSD performance. The NC background is suppressed by two orders of magnitude, expanding JUNO's potential to detect DSNB signals significantly~\cite{JUNO:2022lpc}. We also found the atm-$\nu$ NC with $^{11}$C events are more difficult to reject since these events feature a large $\gamma$ energy fraction from neutron inelastic processes on carbon in the LS. Similar event topologies on oxygen have been observed as background in water Cherenkov detectors. We also studied how much the detector timing response and reconstruction effects degrade the PSD performance. 

The powerful background rejection for atmospheric neutrino NC events provided by the PSD is crucial for the detection of the DSNB in LS experiments. It is important to validate the PSD variable performance with benchmark data in the future. In particular, the dominant NC background channel emits an energetic neutron with $^{11}$C. Therefore, fast neutron sample tagged by the water pool veto detector could be used to constrain the PSD distributions. 
An AmBe neutron calibration source~\cite{JUNO:2020xtj} emits neutrons of several MeVs and an accompanying 4.4 MeV $\gamma$. Events with similar $R_{e\gamma}$ as atm-$\nu$ NC w/ $^{11}$C can be utilized to validate the PSD performance. PSD for atm-$\nu$ NC w/ $^{11}$C events can be evaluated based on events with tagged $^{11}$C decay. 

\section*{Acknowledgements}
The authours would like to thank Guo-Fu Cao and Ze-Yuan Yu for helpful discussions and to thank the JUNO collaboration, especially for providing offline software framework for simulation. We also thank Marta Colomer Molla and Mariangela Settimo for carefully reading the manuscript and valuable comments. This work is supported in part by National Natural Science Foundation of China under Grant Nos. ~12075255, ~11835013, ~12125506, by German Research Foundation under Grant FOR 5519 and by Fundamental Research Funds for the Central Universities (2022048).

\end{document}